\documentclass[fleqn,usenatbib]{mnras}

\usepackage{newtxtext,newtxmath}

\usepackage[T1]{fontenc}
\usepackage{ae,aecompl}

\usepackage{graphicx}	
\usepackage{amsmath}	

\title[Stochastic enrichment of Pop II stars]{The stochastic enrichment of Population II stars}

\author[L. A. Welsh et al.]{
Louise Welsh,$^{1}$\thanks{E-mail: louise.a.welsh@durham.ac.uk}
Ryan Cooke$^{1}$ and
Michele Fumagalli$^{2,1,3}$
\\
$^{1}$Centre for Extragalactic Astronomy, Durham University, South Road, Durham DH1 3LE, UK \\
$^{2}$Dipartimento di Fisica G. Occhialini, Universit\`a degli Studi di Milano Bicocca, Piazza della Scienza 3, 20126 Milano, Italy\\
$^{3}$Institute for Computational Cosmology, Durham University, South Road, Durham DH1 3LE, UK \\}
\date{Accepted XXX. Received YYY; in original form ZZZ}

\pubyear{2020}

\defcitealias{Cayrel2004}{C04}
\defcitealias{Bonifacio2009}{B09}
\defcitealias{Yong2013}{Y13}
\defcitealias{Roederer2014}{R14}
\defcitealias{HegerWoosley2010}{HW10}

\begin{document}
\label{firstpage}
\pagerange{\pageref{firstpage}--\pageref{lastpage}}
\maketitle

\begin{abstract}
We investigate the intrinsic scatter in the chemical abundances of a sample of metal-poor ([Fe/H]~$<-2.5$) Milky Way halo stars. We draw our sample from four historic surveys and focus our attention on the stellar Mg, Ca, Ni, and Fe abundances. Using these elements, we investigate the chemical enrichment of these metal-poor stars using a model of stochastic chemical enrichment. Assuming that these stars have been enriched by the first generation of massive metal-free stars, we consider the mass distribution of the enriching population alongside the stellar mixing and explosion energy of their supernovae. 
For our choice of stellar yields, our model suggests that the most metal-poor stars were enriched, on average, by $\hat{N}_{\star}=5^{+13}_{-3}~(1\sigma)$ Population III stars. This is comparable to the number of enriching stars inferred for the most metal-poor DLAs. Our analysis therefore suggests that some of the lowest mass structures at $z\sim3$ contain the chemical products from $<13~(2\sigma)$ Population III enriched minihaloes. The inferred IMF is consistent with that of a Salpeter distribution and there is a preference towards ejecta from minimally mixed hypernovae. 
However, the estimated enrichment model is sensitive to small changes in the stellar sample. An offset of $\sim0.1$~dex in the [Mg/Ca] abundance is shown to be sensitive to the inferred number of enriching stars. We suggest that this method has the potential to constrain the multiplicity of the first generation of stars, but this will require: (1) a stellar sample whose systematic errors are well understood; and, (2) documented uncertainties associated with nucleosynthetic yields.
\end{abstract}




\section{Introduction}

Before the cosmic dawn, essentially all baryons were comprised of hydrogen and helium. 
The stellar population born from the collapse of this primordial gas transformed their environment irrevocably through the fusion of the first metals (i.e. elements heavier than lithium). As one of the first sources of radiation, these Population III (Pop III) stars were early contributors to the reionisation of the Universe, and, the feedback from these stars influenced the size of the first galaxies \citep{BarkanaLoeb2001,BrommYoshida2011}. This stellar population is also encoded with vital information such as the size and number abundance of the first star-forming structures in the Universe (i.e. the early dark matter minihaloes) \citep{Abel2002, BrommCoppiLarson2002, Bromm2003, Greif2011, Naoz2012}. Studying this stellar population can therefore shed light on parts of the Universe's history that are currently shrouded in mystery. 
\\
\indent In principle, these Population III stars are straight-forward to identify. They are the only stellar population whose atmospheres are expected to be entirely metal-free (at least initially). However, the search for these stars has spanned almost 4 decades and, as of yet, none have been found \citep{Bond1980, Beers1985, Beers1992, Keller2007, Christlieb2008, Aoki2013, Caffau2013, Li2015, Aguado2016, Howes2016, Starkenburg2017, DaCosta2019}. While the first stars have eluded detection, these ongoing surveys have found an ever increasing number of stars that are increasingly deficient in iron. The surviving extremely metal-poor (EMP) stars, whose iron abundances are less than 1/1000 the solar value (i.e. [Fe/H]~$<-3$), are referred to as stellar relics\footnote{Here, and throughout this paper,  [X/Y] denotes the logarithmic number abundance ratio of elements X and Y relative to their solar values X$_{\odot}$ and Y$_{\odot}$, i.e. $[{\rm X / Y}] =  \log_{10}\big( N_{{\rm X}}/N_{{\rm Y}}\big) - \log_{10} \big(N_{{\rm X}}/N_{{\rm Y}}\big)_{\odot}$.}. 
Their surface abundances are thought to be a window to the chemical composition of the gas from which they formed. 
Studying the chemistry of these stars may therefore reveal the properties of the stellar population that preceded them. This approach, termed `stellar archaeology', has become one of the leading observational probes of Population III properties in recent years \citep{Frebel2010}. \\
\indent A key property, yet to be uncovered, is the underlying mass distribution of Population III stars. Current simulations suggest that, as these stars formed in the absence of metals, their typical mass range spanned $\sim 10-100~{\rm M_{\odot}}$ \citep{Clark2008, Stacy2010, Clark2011, Greif2012, Stacy2016}. The Population III initial mass function (IMF) is therefore thought to be distinct from that of later stellar populations. The majority of stars in this expected mass range enrich their environment through core-collapse supernovae (CCSNe). Searches for a Population III signature therefore rely on simulations of stellar evolution, like those of \citet{WoosleyWeaver1995, Umeda2002, Umeda2003, ChieffiLimongi2004, Umeda2005, Tominaga2007, HegerWoosley2010, LimongiChieffi2012}, to compare the chemical abundances expected from Population III CCSNe to those observed in the atmospheres of surviving Population II stars. \\
\indent Carbon-enhanced ([C/Fe]~$>+0.7$) EMP stars that show `normal' relative abundances of neutron-capture elements (i.e. CEMP-no stars) are considered to be the most likely descendants of the first stars \citep{Beers2005, FrebelNorris2015}. It has been suggested that the CEMP-no stars in the metallicity regime $-5<$~[Fe/H]~$<-4$  are the most promising probes of Population III properties \citep{Placco2016}. 
However, only 11 stars are currently known to meet this criteria\footnote{As documented by the Stellar Abundances for Galactic Archaeology (SAGA) database --- an invaluable tool for exploring and compiling stellar samples from existing surveys of  metal-poor stars \citep{Suda2008, Suda2011, Yamada2013, Suda2017}.}. It is also unclear what fraction of Population II stars are likely to present as CEMP-no stars in the local Universe \citep{Ji2015,Hartwig2019}. Thus, searches for a Population III chemical signature generally include EMP stars alongside CEMP-no stars. \\
\indent Cosmological hydrodynamic simulations have suggested that Population III stars likely formed either individually or in small multiples \citep{Greif2010, Stacy2010, Susa2014}. However, traditional comparisons between the observed stellar abundances and the simulated yields have been restricted to two scenarios, either: (1) one progenitor is responsible for the enrichment of a surviving star (e.g. \citealt{Frebel2015, Nordlander2019, Ezzeddine2019}); or (2) the observed abundances can be modelled by the IMF weighted yields from these simulations (e.g. \citealt{HegerWoosley2010, Yong2013, Ishigaki2018}). Though, the use of semi-analytic models has enabled the consideration of multiple enriching progenitors \citep{Karlsson2005a, Karlsson2005b, Hartwig2018Oct}. \\
\indent In this paper, we use a novel approach to analyse the chemistry of the most metal-poor stars, and use this tool to infer the number of massive Population III stars that have enriched the surviving metal-poor stellar relics. Previously, we have applied this tool to investigate the enrichment of the \emph{most} metal-poor damped Lyman-$\alpha$ systems (DLAs) \citep{Welsh2019, Welsh2020}. We now extend this work to investigate the stochastic enrichment of a sample of metal-poor Milky Way halo stars using their measured [Mg/Ca] and [Ni/Fe] abundances.
While not the subject of this work, we note that the potential for inhomogeneous metal-mixing at the sites of Population II star formation is an additional complication that is an interesting avenue for further investigations \citep{Salvadori2010, Sarmento2017, Hartwig2018Aug, Tarumi2020}. \\
\indent Our paper is organised as follows. Section~\ref{sec:data} describes the observational sample used in this paper. Section~\ref{sec:scatter} motivates the use of a stochastic enrichment model. In Section~\ref{sec:model} we outline this model and apply it to the observational data. The results are discussed in Section~\ref{sec:disc} before drawing overall conclusions in Section~\ref{sec:conc}.

\begin{figure}
	\includegraphics[width=\columnwidth]{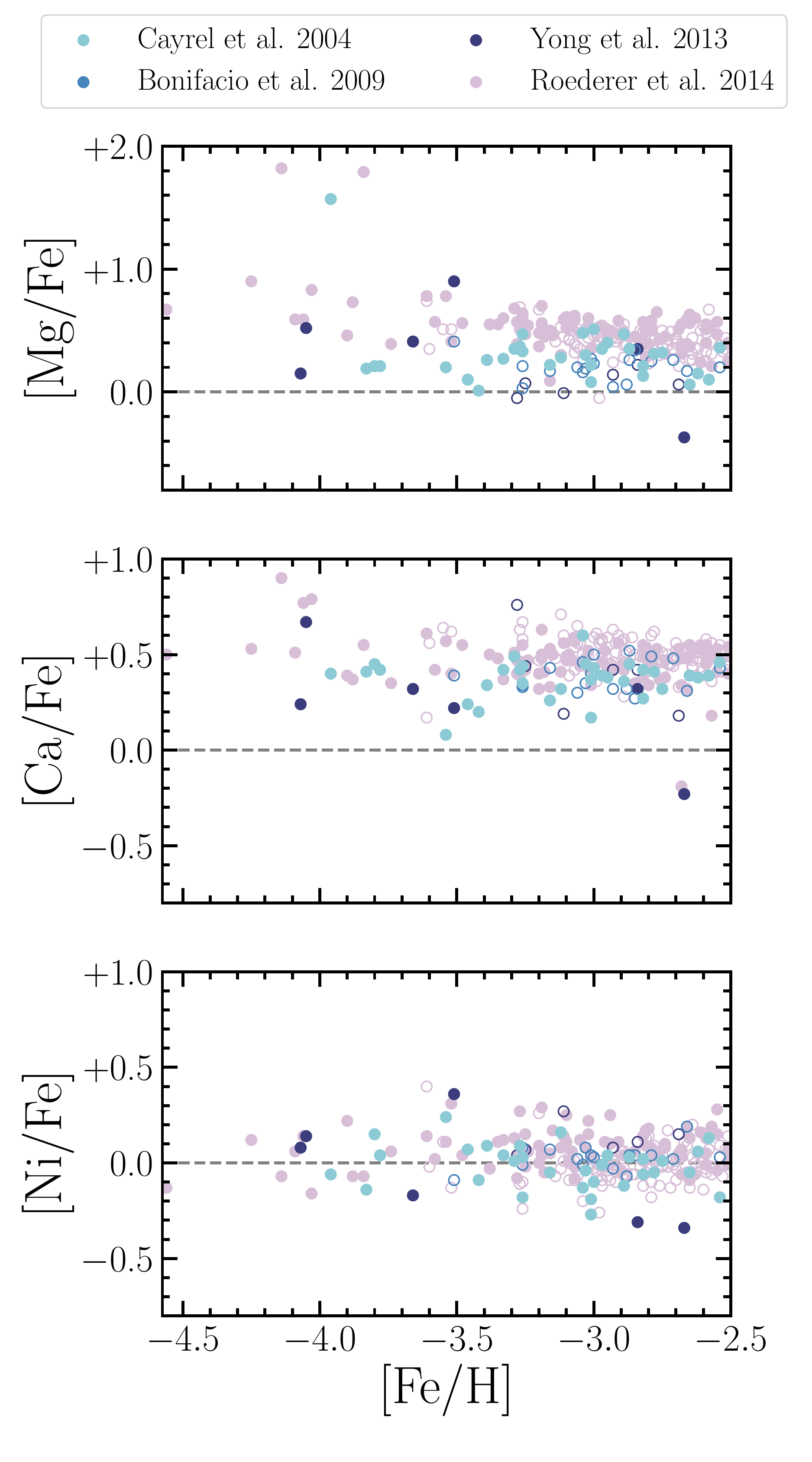}
    \caption{From top to bottom, the successive panels show the measured [Mg/Fe], [Ca/Fe], and [Ni/Fe] abundances of our stellar sample as a function of their [Fe/H] abundances. The colour of the marker indicates the source of the data (defined in the legend and used throughout this paper). The fill of the marker indicates the evolutionary stage of the star --- giants are shown by filled circles while non-giants are shown by hollow circles. The horizontal dashed line indicates the solar value. Note the different y-axis scale used in the top panel.
    }
    \label{fig:raw_data}
\end{figure}

\section{Data}
\label{sec:data}
The stellar abundances considered in this work are a compilation of four sources. Specifically, that of: \citeauthor{Cayrel2004} \citepalias[2004; hereafter][]{Cayrel2004}, \citeauthor{Bonifacio2009} \citepalias[2009; hereafter][]{Bonifacio2009}, \citeauthor{Yong2013} \citepalias[2013; hereafter][]{Yong2013}, and, \citeauthor{Roederer2014} \citepalias[2014; hereafter][]{Roederer2014}. \citetalias{Cayrel2004} and \citetalias{Bonifacio2009} are part of the \textit{First Stars} series. 
Note that we only consider the programme stars reported by \citetalias{Yong2013} (and not the literature stars used in their analysis). A summary of these data can be found in Table~\ref{tab:surveys} where programme size indicates the number of stars reported in the original works, sample size indicates the number of objects used in our analysis, and $\sigma_{\rm X}$ indicates the median error associated with the reported [X/Fe] values.
\begin{table}
	\centering
	\caption{Summary of surveys that we consider in this work.}
	\label{tab:surveys}
\begin{tabular}{lcccccc}
\hline
survey & \begin{tabular}[c]{@{}l@{}}programme\\ size\end{tabular} & \begin{tabular}[c]{@{}l@{}}sample \\ size\end{tabular} & N$_{\rm giants}$ & \multicolumn{1}{l}{$\sigma_{\rm Mg}$} & \multicolumn{1}{l}{$\sigma_{\rm Ca}$} & \multicolumn{1}{l}{$\sigma_{\rm Ni}$} \\ 
\hline
\citetalias{Cayrel2004}   & 35     & 32      & 32   &  0.12 &  0.11 & 0.06   \\
\citetalias{Bonifacio2009}   & 19          & 18     & 0      &  0.06 &  0.11 & 0.07         \\
\citetalias{Yong2013}   & 38    & 12      & 6     &  0.11 &  0.12 & 0.15             \\
\citetalias{Roederer2014}    & 313      & 188   & 92       &  0.11 &  0.15 & 0.17       \\                      
\hline
\end{tabular}
\end{table}
Across all samples there are 11 metals in common\footnote{These elements include: C, Mg, Si, Ca, Sc, Ti, Cr, Mn, Fe, Co, and Ni.}. In our paper, we restrict our analysis to the Mg, Ca, Fe, and Ni relative abundances of these samples. The choice of elements is primarily driven by the perceived confidence in both the observed abundances and the simulated yields. Furthermore, the abundances of these elements are amongst those most commonly reported throughout the chosen stellar samples. Additionally, they are sensitive to the properties (e.g. mass, metallicity, stellar mixing etc.) of the stars that synthesised these elements. Finally, as discussed in Section~\ref{sec:model}, our modelling technique is computationally expensive, so we are currently limited to selecting only a small number of the most reliable elements.\\ 
\indent The abundances adopted in our work have been computed under the assumption of local thermodynamic equilibrium (LTE) using 1D model atmospheres. These models do not capture possible spatial inhomogenities. Mg, Ca, and Ni abundances are often determined from the spectral features of neutral species; these lines are generally thought to be more susceptible to non-LTE processes than those of ionised species \citep{Asplund2005}. 
We do not apply non-LTE corrections to these data, but the impact of this decision is discussed in Section~\ref{sec:nlte}. Similarly, Fe\,{\sc i} lines are known to be affected by overionisation \citep{Thevenin1999}. This can in-turn impact the estimated surface gravity of the star. These departures from LTE are often accounted for through comparisons of the Fe\,{\sc i} and Fe\,{\sc ii} abundances. We note that this correction may be imperfect as Fe\,{\sc ii} lines may also form in regions that depart from LTE. \\
\begin{figure*}
	\includegraphics[width=\textwidth]{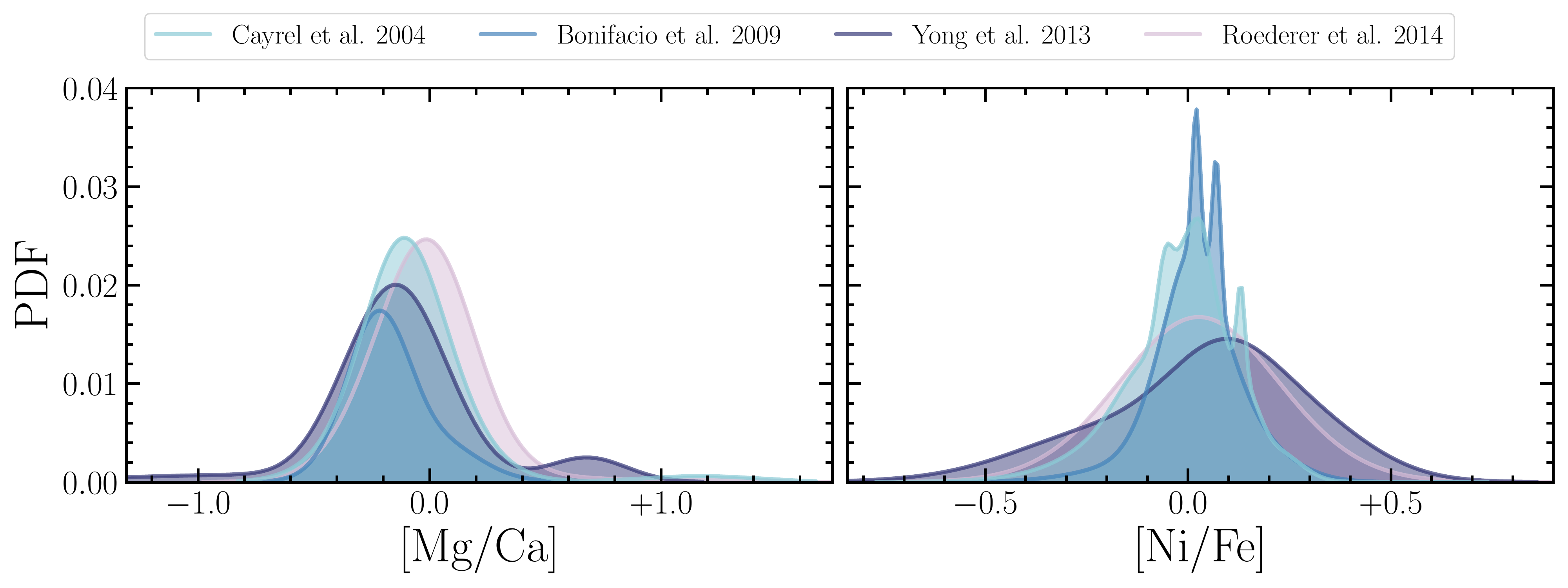}
    \caption{Generalised histograms showing the [Mg/Ca] (left) and [Ni/Fe] (right) abundance distributions of the individual samples. The colour indicates of the source of the data. The chemical abundance of each object has been treated as a Gaussian with a standard deviation given by the observational error. Considering the abundance ratios of elements close to one another in atomic number minimises the differences seen between the samples.}
    \label{fig:gen_hist}
\end{figure*}
\indent We apply a metallicity cut at [Fe/H]~$<-2.5$. This ensures our sample of metal-poor halo stars is of comparable metallicity to the sample of DLAs analysed by \citet{Welsh2019}. This criteria is more lax than the limit commonly imposed when investigating Population II chemical enrichment; these studies often reserve their analysis for EMP stars with [Fe/H]~$<-3$. These stars are considerably less numerous and are often analysed individually; including these stars, from additional sources, would risk biasing our sample towards the \emph{most} chemically peculiar stars currently known. As summarised in Table~\ref{tab:surveys}, we consider 32 stars from \citetalias{Cayrel2004}, 18 stars from \citetalias{Bonifacio2009}, 12 stars from \citetalias{Yong2013}, and 188 stars from \citetalias{Roederer2014}. 
Figure \ref{fig:raw_data} shows the abundances of these survey stars as a function of their [Fe/H] abundance. 
Note that to be included in our sample, a star must be sufficiently metal poor ([Fe/H]~$<-2.5$) and have bounded [Mg/Fe], [Ca/Fe], and [Ni/Fe] abundances. 
This sample contains stars in various stages of their evolution. We make no initial distinction between the abundances of giants and non-giants --- though, they can be distinguished in Figure \ref{fig:raw_data} through the fill of the markers. Note that below [Fe/H]~$<-3.5$, our sample almost exclusively contains giant stars (which are known to be more susceptible than dwarfs to non-LTE processes). \\
\indent The adopted metallicities, [Fe/H], are determined solely from Fe\,{\sc i} abundances. The relative abundance ratio between any two elements is determined by considering the abundances of species in the same ionisation state. In these works, Mg is determined using the spectral features of neutral species, therefore [Mg/Fe] is given by  [Mg\,{\sc i}/Fe\,{\sc i}], together with its associated error.\footnote{Note, in the case of \citetalias{Cayrel2004}, the errors of [X\,{\sc i}/Fe\,{\sc i}] are given by the errors reported for [X/Fe].} 
The solar values adopted in this paper are taken from \cite{Asplund} and are shown in Table~\ref{tab:solar_vals} alongside those adopted by the original sources. Note that all of the measured relative abundances and chemical yields that are used in this work have been registered onto the same solar abundance scale (see second column of Table~\ref{tab:solar_vals}).\\
\begin{table}
	\centering
	\caption{Solar abundances adopted in this analysis and those used in sample sources.}
	\label{tab:solar_vals}
	\begin{tabular}{ccccc} 
		\hline
		Element & This Work  & \citetalias{Cayrel2004} & \citetalias{Bonifacio2009} & \citetalias{Yong2013} $/$ \citetalias{Roederer2014} \\
		\hline
		Mg & 7.56 & 7.58 & 7.58 & 7.60 \\
		Ca & 6.29 & 6.36 & 6.36 & 6.34 \\
		Fe & 7.47 & 7.50 & 7.51 & 7.50 \\
		Ni & 6.21 & 6.25 & 6.25 & 6.22 \\
		\hline
	\end{tabular}
\end{table}
\indent To investigate the properties of Population III stars using the chemical abundances of Population II stars, we require a homogeneous stellar sample whose photospheric abundances are not dominated by systematic effects. As highlighted in Figure~\ref{fig:raw_data}, the \citetalias{Roederer2014} data is known to show an elevated [Mg/Fe] abundance relative to the other samples. The origin of this offset, as found by \citetalias{Roederer2014}, is due to the way that the effective temperature $T_{\rm eff}$ is determined. 
\citetalias{Cayrel2004}, \citetalias{Bonifacio2009}, and \citetalias{Yong2013} each utilise a combination of photometric and spectroscopic information to determine $T_{\rm eff}$. \citetalias{Roederer2014} exclusively consider spectroscopic data in their determination of $T_{\rm eff}$. 
We find that this offset between the \citetalias{Roederer2014} data and the other samples can be minimised when considering the abundance ratios of elements close to one another in atomic number --- in this case [Mg/Ca] and [Ni/Fe]. The generalised histograms of these abundance ratios for each sample are shown in Figure~\ref{fig:gen_hist}. 
When comparing the abundances of these stars to the yields of Population III SNe, we therefore draw parallels between the observed and simulated [Mg/Ca] and [Ni/Fe] abundances\footnote{Note that the distributions shown in Figure~\ref{fig:gen_hist} are almost identical for giants and non-giants.}. This allows us to consider the stars from all four surveys simultaneously and take advantage of the large dataset afforded by \citetalias{Roederer2014}. 
We note that there are a number of stars that appear in multiple surveys. For these objects, we adopt the abundances derived from data with the superior spectral resolution. The numbers quoted in Table~\ref{tab:surveys} represent the data after the removal of duplicates; there are 250 unique stars in this sample. \\
\indent Finally, we note that the stars comprising the \textit{First Stars} series (\citetalias{Cayrel2004}, \citetalias{Bonifacio2009}) are thought to be kinematically associated with either the Gaia Sausage-Enceladus satellite or \textit{in-situ} star formation \citep{DiMatteo2020}. In the context of our work, it is important that we focus on only the \textit{most} metal-poor stars; at somewhat higher metallicity, the signature of Pop III stars will be increasingly washed out by the early star formation history of the dwarf galaxy where the Pop II star was born. Thus, provided our stellar sample are purely enriched by Pop III stars, the origin of the star is not critical to our analysis.

\section{Intrinsic scatter}
\label{sec:scatter}
\begin{figure*}
    \centering
    \includegraphics[width=\textwidth]{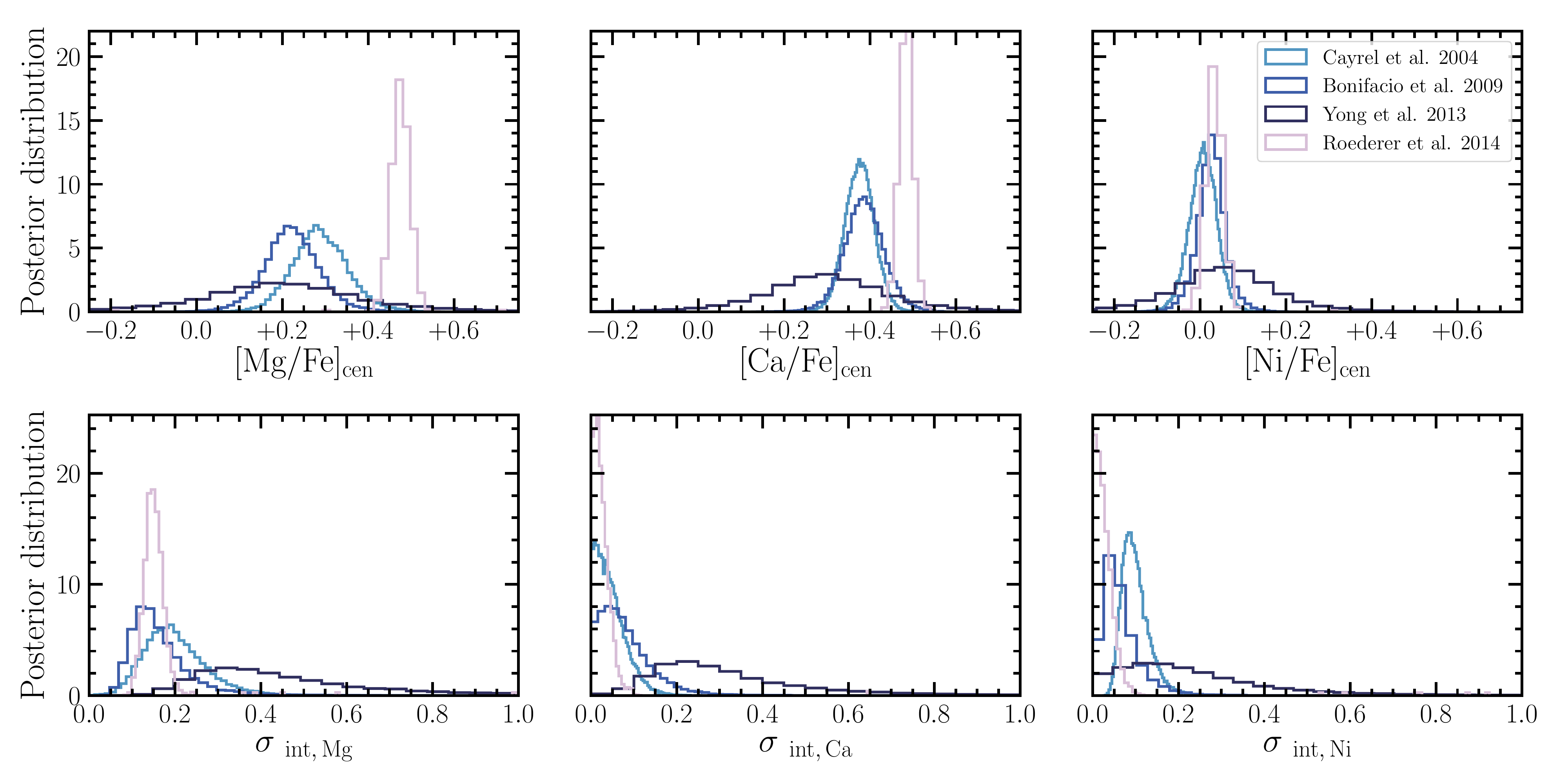}
    \caption{The top row shows the posterior distributions of the central [Mg/Fe] (left), [Ca/Fe] (centre), and [Ni/Fe] (right) abundance ratios estimated for each survey independently. The bottom row shows the posterior distributions of the intrinsic scatter associated with these central values. The colour of the histogram indicates the source of the data (as indicated in the legend). 
    }
    \label{fig:all_auth_scatter}
\end{figure*}
\begin{figure*}
	\includegraphics[width=0.8\textwidth]{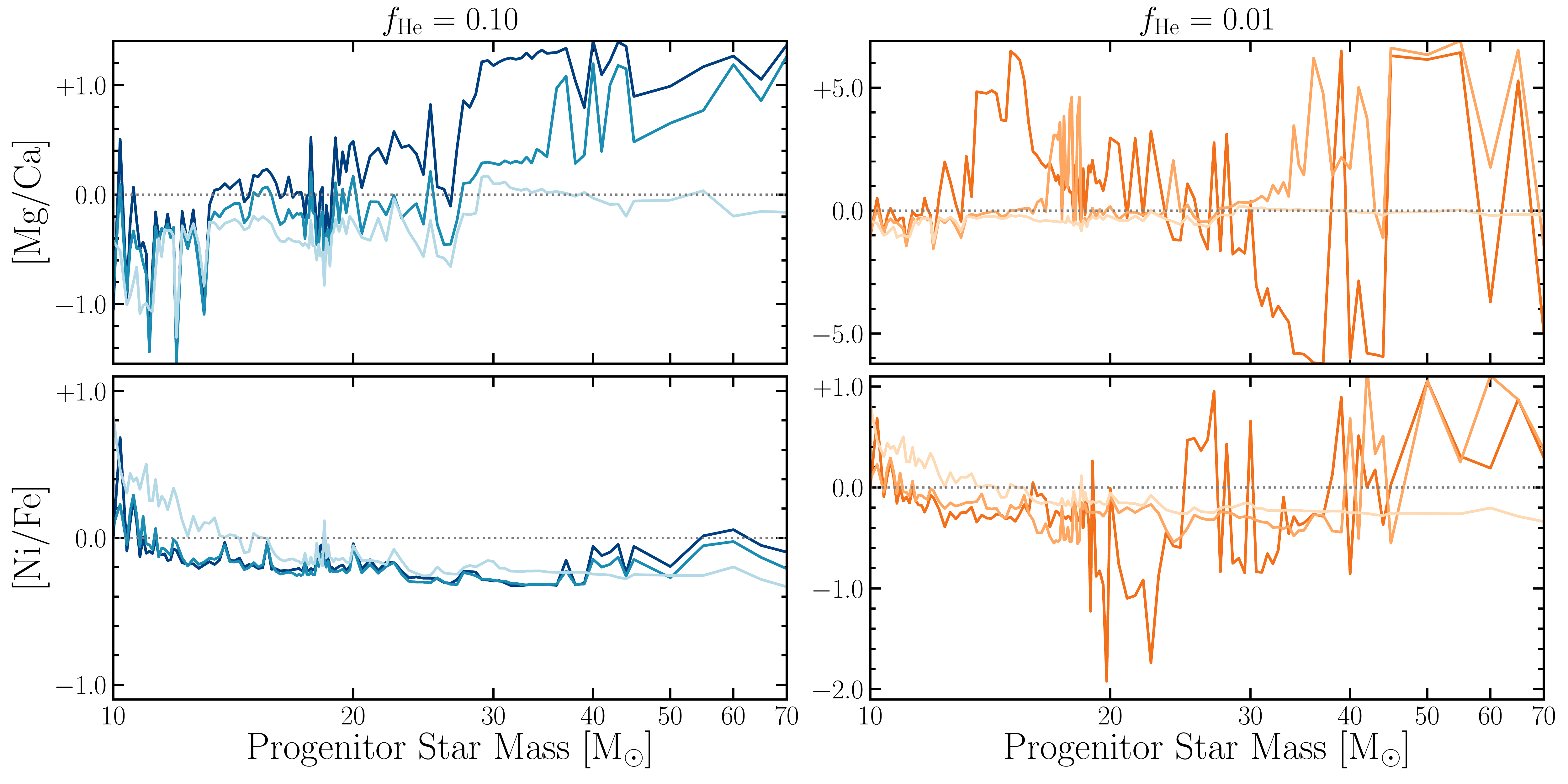}
    \caption{[Mg/Ca] and [Ni/Fe] yields as a function of the progenitor star mass for a range of stellar properties. As indicated by the title, the two left-hand panels show the yields associated with stars that undergo the fiducial amount of mixing between stellar layers (10 per cent of the He core size). The progressively lighter blue lines correspond to progressively more energetic explosions. Specifically, dark to light shades highlight a 0.9, 1.8, and 10~B explosion, respectively. The orange lines in the right-hand panels highlight the variability of the yields when a star experiences minimal mixing between the stellar layers (1 per cent of the He core size). The progressively lighter orange lines correspond to the same change in explosion energy. Finally, the horizontal dotted line indicates the solar abundance of these abundance ratios.}
    \label{fig:yields_2x2_exp}
\end{figure*}
Before we model the chemical enrichment of these data in detail, we first motivate the importance of employing a stochastic chemical enrichment model given the present sample of stars. Treating each stellar survey independently, we have investigated if the reported abundances show significant \emph{intrinsic} scatter that cannot be explained by the observational errors. To quantify this additional scatter, we model each abundance ratio as a Gaussian centred about [X/Fe]$_{\rm cent}$ whose error is given by the expression:
\begin{equation}
\label{eqn:sig}
\sigma_{\rm tot}^{2} = \sigma_{\rm obs}^{2} + \sigma_{\rm int}^{2} \,.
\end{equation}
Thus, the dispersion of the data is given by the observed error $\sigma_{\rm obs}$ and an additional intrinsic component $\sigma_{\rm int} $ added in quadrature. We therefore consider two model parameters ([X/Fe]$_{\rm cent}$ and $\sigma_{\rm int}$) that we determine independently for each abundance ratio and each sample. \\
\indent When looking at the [Mg/Fe], [Ca/Fe], and [Ni/Fe] abundances of the individual surveys, we have consistently found a non-zero intrinsic scatter associated with the [Mg/Fe] abundances of all the stellar samples. This intrinsic scatter is estimated using a Markov Chain Monte Carlo (MCMC) procedure that simultaneously estimates the central values [Mg/Fe]$_{\rm cent}$, [Ca/Fe]$_{\rm cent}$, and [Ni/Fe]$_{\rm cent}$ of a sample, alongside their associated additional error components $\sigma_{\rm int, Mg}$, $\sigma_{\rm int, Ca}$, and $\sigma_{\rm int, Ni}$. For details of this calculation and an example of the converged analysis of the \citetalias{Cayrel2004} data, see Appendix~\ref{sec:append}. Figure \ref{fig:all_auth_scatter} shows the resulting posterior distributions of this analysis for all stellar samples used in this work. 
The bottom left panel of this figure shows the posterior distribution of the intrinsic scatter present in the [Mg/Fe] data. This intrinsic scatter implies that either: (1) there is a consistent intrinsic scatter associated with the [Mg/Fe] abundances of the stars in every sample; or (2) the [Mg/Fe] abundance errors are all consistently underestimated by $\sim0.15$~dex. In this paper, we consider the first possibility, but note that there may also be some non-physical origin of this scatter.  
For example, the impact of assuming LTE may affect stars of a given metallicity, temperature, and/or surface gravity more significantly than others. \\ 
\indent We have repeated this analysis after removing the peculiar stars with [Mg/Fe]$>+1$ (see the top panel of Figure~\ref{fig:raw_data}). This modification affects both the \citetalias{Cayrel2004} and \citetalias{Roederer2014} data. We find that the estimated intrinsic scatter is reduced across both samples; however, both the revised \citetalias{Cayrel2004} and revised \citetalias{Roederer2014} samples independently support a statistically significant deviation ($1.5\sigma$) from zero intrinsic scatter.
We have also investigated whether this scatter is dependent on the metallicity of the star. Focusing on the \citetalias{Cayrel2004} sample, we found that the enhanced [Mg/Fe] abundance of CS~$22949-037$ is difficult to replicate alongside the other data. If it is removed from the sample, we find no strong dependence with metallicity. The result is mirrored when considering the \citetalias{Roederer2014} data. The \citetalias{Bonifacio2009} and \citetalias{Yong2013} samples are deemed too small to reliably capture any relationship between the intrinsic scatter and the metallicity of a star. \\
\indent We emphasise that, irrespective of trends with metallicity,  the intrinsic scatter in [Mg/Fe] is non-negligible for each sample when treated independently. The scatter in this data may highlight the inhomogeneous nature of the environments within which these stars formed. To understand the chemical abundances of these objects it is therefore necessary to consider a stochastic chemical enrichment model. \\\begin{figure*}
	\includegraphics[width=0.8\textwidth]{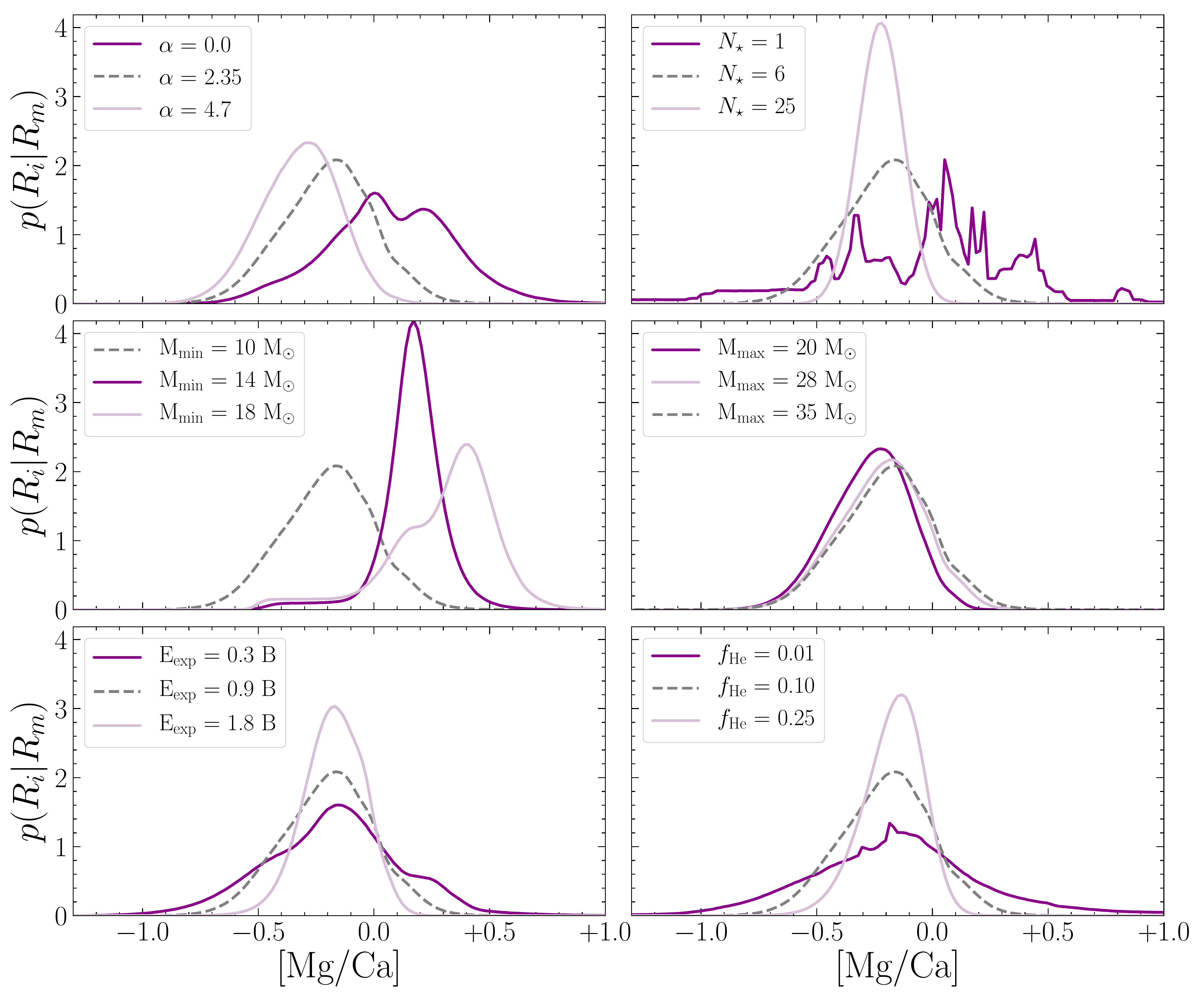}
    \caption{Probability distribution of the intrinsic [Mg/Ca] ratio that would be expected for a range of underlying enrichment models. From top left to bottom right, the successive panels correspond to changing the slope, number of enriching stars, minimum mass, maximum mass, explosion energy, and degree of mixing. 
    Unless otherwise stated in the legend, the model parameters of these distributions are $\alpha = 2.35$, $N_{\star} = 6$, $M_{\rm min} = 10$ M$_{\odot}$, $M_{\rm max} = 35$ M$_{\odot}$, $E_{\rm exp} = 0.9$ B, and $f_{\rm He}=10$~per cent (displayed as the grey dashed line in all panels as a point of comparison).}
    \label{fig:PDF_1D}
\end{figure*}
\section{Stochastic Enrichment}
\label{sec:model}

Using a model of stochastic chemical enrichment in combination with the yields from simulations of stellar evolution, we can test if the observed chemical abundances of Population II stars are consistent with enrichment by Population III SNe. 
 This model assumes that the number of Population III stars responsible for enriching a given environment is given by the integral:
\begin{equation}
\label{eqn:N}
    N_{\star}  = \int^{M_{\rm max}}_{M_{\rm min}}kM^{-\alpha}{\rm d}M
\end{equation}
where $M$ is the mass of the Population III star, $\alpha$ controls the slope of the power-law, and $k$ is a multiplicative constant dependent on the number of stars forming within the mass range bounded by $M_{\rm min}$ and $M_{\rm max}$. In addition to the mass distribution of the enriching stars, we consider the typical explosion energy of the Population III SNe as well as the degree of mixing between the stellar layers. To determine the abundances expected from a given model, we rely on simulations of stellar evolution. Our fiducial choice of nucleosynthetic yields, described below, are those of  \citeauthor{HegerWoosley2010} \citepalias[2010; hereafter][]{HegerWoosley2010}. \\
\indent The \citetalias{HegerWoosley2010} simulations calculate the yields of massive Population III stars that end their lives as Type II SNe. This simulation suite reports a grid of chemical yields as a function of: (1) the progenitor star mass; (2) the mixing between the stellar layers during the explosion, and; (3) the kinetic energy of the SNe ejecta at infinity. These yields have been calculated using non-rotating, 1D models under the assumption of spherical symmetry. They do not account for mass loss or magnetic fields. 
The Rayleigh-Taylor instabilities that aide mixing between the stellar layers cannot be captured by 1D models. In these simulations mixing is achieved by moving a boxcar of width $\Delta M$ through the star, typically four times. $\Delta M$ is described as a fraction of the He core size $f_{\rm He}$. The boxcar width that reproduces the hard X-ray and optical lightcurves of SN 1987A is 10 per cent of the He core size ($f_{\rm He}=0.100$). The eventual collapse of the progenitor, and the subsequent SN explosion, is simulated by depositing momentum at the base of the oxygen burning shell. The strength of this explosion is parameterized by the kinetic energy of the ejecta at infinity. 
The typical explosion energy of a SN is $E_{\rm exp} = 1$~B (where $1~{\rm B} = 10^{51}~{\rm ergs}$). 
The \citetalias{HegerWoosley2010} yields have been calculated for 16,800 combinations of these three parameters (progenitor mass, stellar mixing, and explosion energy). \\
\indent Figure~\ref{fig:yields_2x2_exp} shows the yields of [Mg/Ca] and [Ni/Fe] as a function of progenitor star mass for a range of explosion energies. The left and right set of panels show the simulated yields for two mixing prescriptions. Those on the left adopt the value recommended by \citetalias{HegerWoosley2010} (10 per cent of the He core size). Those on the right indicate the yields of SNe that undergo minimal mixing between stellar layers (1 per cent of the He core size). 
From this figure it is clear that the yields of both [Mg/Ca] and [Ni/Fe] are highly variable for low values of the SNe explosion energy and/or the mixing width. There is a more consistent relationship with progenitor mass when $f_{\rm He} = 0.1$. In this case, there is a general trend of increasing [Mg/Ca] with increasing progenitor mass, while [Ni/Fe] shows no strong evolution across the considered mass range. 
\subsection{Stochastic sampling}
\label{sec:sampling}

To determine how the simulated Population III yields compare to the observed abundances of stellar relics, it is necessary to consider how these objects may have been chemically enriched. In their respective minihaloes, the first stars are thought to form either individually or in small multiples. Thus, the surviving Population II stars may have been enriched by the chemical products of multiple SNe. The progenitors of these SNe formed obeying an underlying mass distribution. Given the small number of Population III stars forming in each minihalo, this mass distribution would have necessarily been stochastically sampled. The relative abundances of the stellar population enriched by Population III SNe may therefore show an intrinsic spread. \\
\indent For any given enrichment model, we would like to calculate the probability of observing each star in our sample. The probability of observing a given abundance pattern (e.g. [Mg/Ca] and [Ni/Fe]) is
\begin{equation}
\label{eqn:prob}
    p_{n} \big( R_{o} | R_{m} \big)  = \int  p \big( R_{o} | R_{i} \big)  p \big( R_{i} | R_{m} \big) {\rm d} R_{i}  \; .
\end{equation}
The first term of this integral describes the probability of \emph{observing} a given abundance pattern ($R_{o}$) given the intrinsic abundance ratios of the system, $R_{i}$. In other words, this variable describes how close the observed measurement is to the true value. We model each abundance ratio by a Gaussian whose spread is given by the observational error. The second term of this integral describes the probability of obtaining an intrinsic abundance pattern given the stochastic enrichment model defined by Equation \ref{eqn:N}. \\
\indent To determine the expected distribution of relative abundances given an enrichment model (i.e. $p(R_{i}|R_{m})$), we construct a grid of IMF model parameters, and sample each grid point $10^{3}$ times. For each iteration, we use the sampled stellar masses to determine the yields of the associated SNe. It is assumed that the SNe ejecta are well-mixed. The resulting number abundance ratio of [Mg/Ca], for example, is therefore given by the total yield of Mg relative to the total yield of Ca from all of the sampled stars. 
Using this Monte Carlo sampling technique, we can build an \emph{N}-dimensional probability density function of the expected yields, where \emph{N} is the number of abundance ratios under consideration. In this work we consider the [Mg/Ca] and [Ni/Fe] abundance ratios (thus, \emph{N}=2).  \\
\indent In this work, $p (R_{i} | R_{m})$ describes the joint probability of simultaneously producing any given combination of [Mg/Ca] and [Ni/Fe]. Figure~\ref{fig:PDF_1D} illustrates how the different enrichment model parameters influence the expected distribution of [Mg/Ca]. The successive panels correspond to changing the slope, number of enriching stars, minimum mass, maximum mass, explosion energy, and stellar mixing, respectively. The fiducial model parameters used in Figure~\ref{fig:PDF_1D} (grey-dashed curves) are: $\alpha = 2.35$, $N_{\star} = 6$, $M_{\rm min} = 10~{\rm M_{\odot}}$, $M_{\rm max} = 35~{\rm M_{\odot}}$, $E_{\rm exp}=0.9~{\rm B}$, and $f_{\rm He}=10$~per cent. 
From this figure it is clear that the expected distributions of intrinsic abundances are sensitive to the average number of Pop III stars that enriched the second generation of stars. Thus, given an appropriate sample of observed stellar abundances, this model can be used to estimate the average number of Population III SNe that have chemically enriched the surviving Population II stars. Under the assumption that each surviving star is enriched by the well-mixed SNe ejecta from one minihalo, this analysis can be used to gauge the multiplicity of the first stars.

\subsection{Likelihood analysis}
\label{sec:ana}
\indent The likelihood of an enrichment model is given by:
\begin{equation}
\label{eqn:like}
        \mathcal{L} = \prod_{n} p_{n} ( R_{o} | R_{m} ) \; ,
\end{equation}
To estimate the enrichment model parameters that provide the best fit to the observed abundances, we utilise the {\sc emcee} software package \citep{EMCEE} to conduct an MCMC likelihood analysis.
In this analysis, we adopt uniform priors across all of the model parameters bounded by: 
\begin{eqnarray*}
 1 \leq &N_{\star}& \leq 100 \; , \\
 -5 \leq &\alpha &\leq 5 \; , \\
20\leq &M_{\rm max}/{\rm M_{\odot}}& \leq 70 \; , \\
 0.3 \leq &E_{\rm exp}/10^{51}{\rm erg}& \leq 10  \; , \\
  0 \leq &f_{\rm He}& \leq 0.25 \; .
\end{eqnarray*}
\begin{figure*}
	\includegraphics[width=\textwidth]{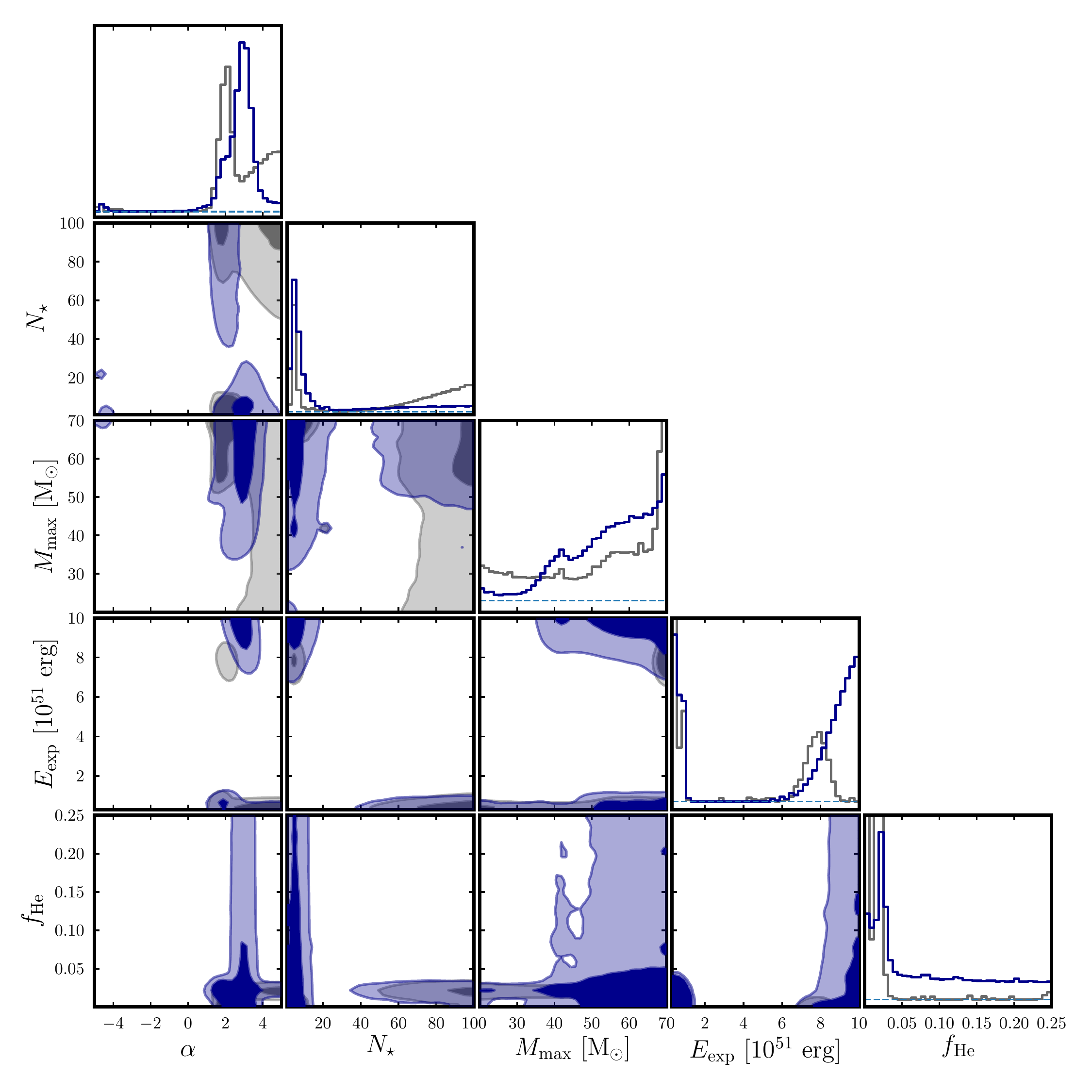}
    \caption{The marginalised maximum likelihood distributions of our stochastic chemical enrichment model parameters (main diagonal), and their associated 2D projections, given the abundances of a sample of metal-poor MW halo stars. The dark and light contours show the $68\%$ and $95\%$ confidence regions of these projections, respectively. The horizontal blue dashed lines in the diagonal panels mark where the individual parameter likelihood distributions fall to zero. The grey distributions correspond to the simultaneous analysis of the stars from \protect\citetalias{Cayrel2004}, \protect\citetalias{Bonifacio2009}, \protect\citetalias{Yong2013}, and \protect\citetalias{Roederer2014}. The blue distributions are the result of removing the \protect\citetalias{Roederer2014} data from the sample.}
    \label{fig:corner}
\end{figure*}
These boundary conditions are chosen to cover the physically motivated parameter space, given our assumptions about Population III star formation. The number of massive stars that chemically enrich their environment via CCSNe is expected to be small. The slope of the power-law that dictates Population III star formation  is still unknown, we therefore have parameterised the IMF of massive stars ($M > 10~M_{\odot}$) as a power-law IMF, which is consistent with the shape found in the local Universe. The slope of this power-law for local star formation is given by $\alpha = 2.35$ (\citealt{Salpeter1955}; see \citealt{Bastain2010} for a review). The minimum mass of the enriching stars is fixed at $M_{\rm min} = 10~{\rm M_{\odot}}$ as it is assumed that all stars above this mass limit are capable of undergoing core-collapse. The upper bound on the mass of the enriching stars coincides instead with the onset of pulsational pair instability SNe \citep{Woosley2017}. 
The boundary conditions on both the average SN explosion energy and mixing prescription cover all of the values explored in \citetalias{HegerWoosley2010}. These simulations calculated the SNe yields for discrete combinations of $M$, $E_{\rm exp}$, and $f_{\rm He}$; we linearly interpolate over this 3D parameter space for our analysis. \\
Similar to the approach from \citet{Welsh2019, Welsh2020}, we begin our analysis with 400 randomly initialised walkers. These walkers each take 10 000 steps to converge on the stable posterior distributions shown in Figure~\ref{fig:corner}. 
\begin{figure*}
	\includegraphics[width=\textwidth]{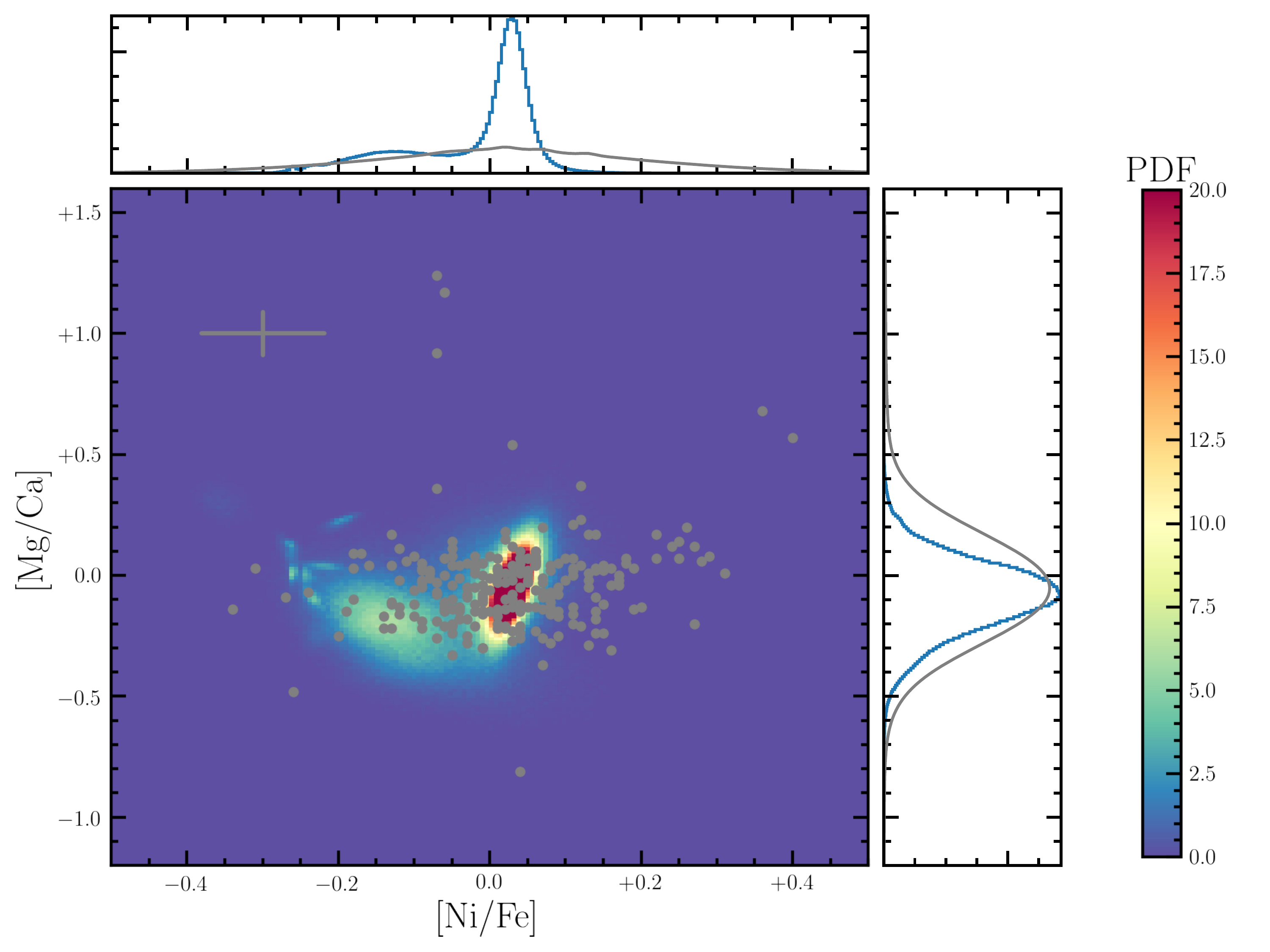}
    \caption{[Mg/Ca] and [Ni/Fe] data of all of the samples with the median error plotted in the top left corner. The background contours highlight the PDF of the expected abundances given our inferred chemical enrichment model. The blue histograms in the top and left panels show the 1D projections of the expected [Ni/Fe] and [Mg/Ca] abundances, respectively, given our inferred enrichment model. The grey curves show the generalised histograms of the data used in our analysis.}
    \label{fig:model_v_data_all}
\end{figure*}
The grey contours highlight the result of considering all of the stars in our chosen sample. We find that the mass distribution of the enriching Population III stars is broadly consistent with a Salpeter ($\alpha = 2.35$) IMF; however, the bottom-heavy tail of the distribution is poorly constrained. Similarly, the maximum mass of the enriching stars is largely unconstrained, while showing a slight preference towards higher values. This analysis suggests that the enriching progenitors have experienced minimal mixing between stellar layers ($f_{\rm He}\sim0.03$). The remaining parameters, $N_{\star}$ and $E_{\rm exp}$, show two possible scenarios. If $N_{\star}<20$ then the yields of hypernovae ($E_{\rm exp}\sim8$~B) are most suitable. Alternatively, if  $N_{\star}>30$, we find that these Population II stars are best modelled with the yields of weak  ($E_{\rm exp}\sim0.3$~B) Population III SNe. This latter scenario is consistent with the result of \citetalias{HegerWoosley2010} (see their figure~12). 
If we repeat this analysis after removing the \citetalias{Roederer2014} data (see the blue contours in Figure~\ref{fig:corner}), these two scenarios persist. However, there is a clear preference towards the low $N_{\star}$ scenario that, in this case, also coincides with an unconstrained mixing prescription. Removing the \citetalias{Roederer2014} data reduces the allowed parameter space of our model, even though the observed distributions of [Mg/Ca] and [Ni/Fe] of each independent survey are broadly consistent (see Figure~\ref{fig:gen_hist}). This highlights that the estimated model parameters are sensitive to small differences in the observationally measured abundances. 
Before we investigate the potential origins of this difference, we note that repeating our analysis using only the abundances of the 130 giants in our sample (defined as those with $\log g < 3$) has a negligible impact on our parameter estimates. Similarly, removing potentially peculiar stars with elevated Mg ([Mg/Fe]~$>+1$) does not change the parameter estimates. This is not surprising, as the majority of the stars in our sample have [Mg/Fe]~$< +1$ (see Figures~\ref{fig:raw_data} and \ref{fig:gen_hist}).

\subsection{Maximum likelihood results}
\begin{figure*}
	\includegraphics[width=\textwidth]{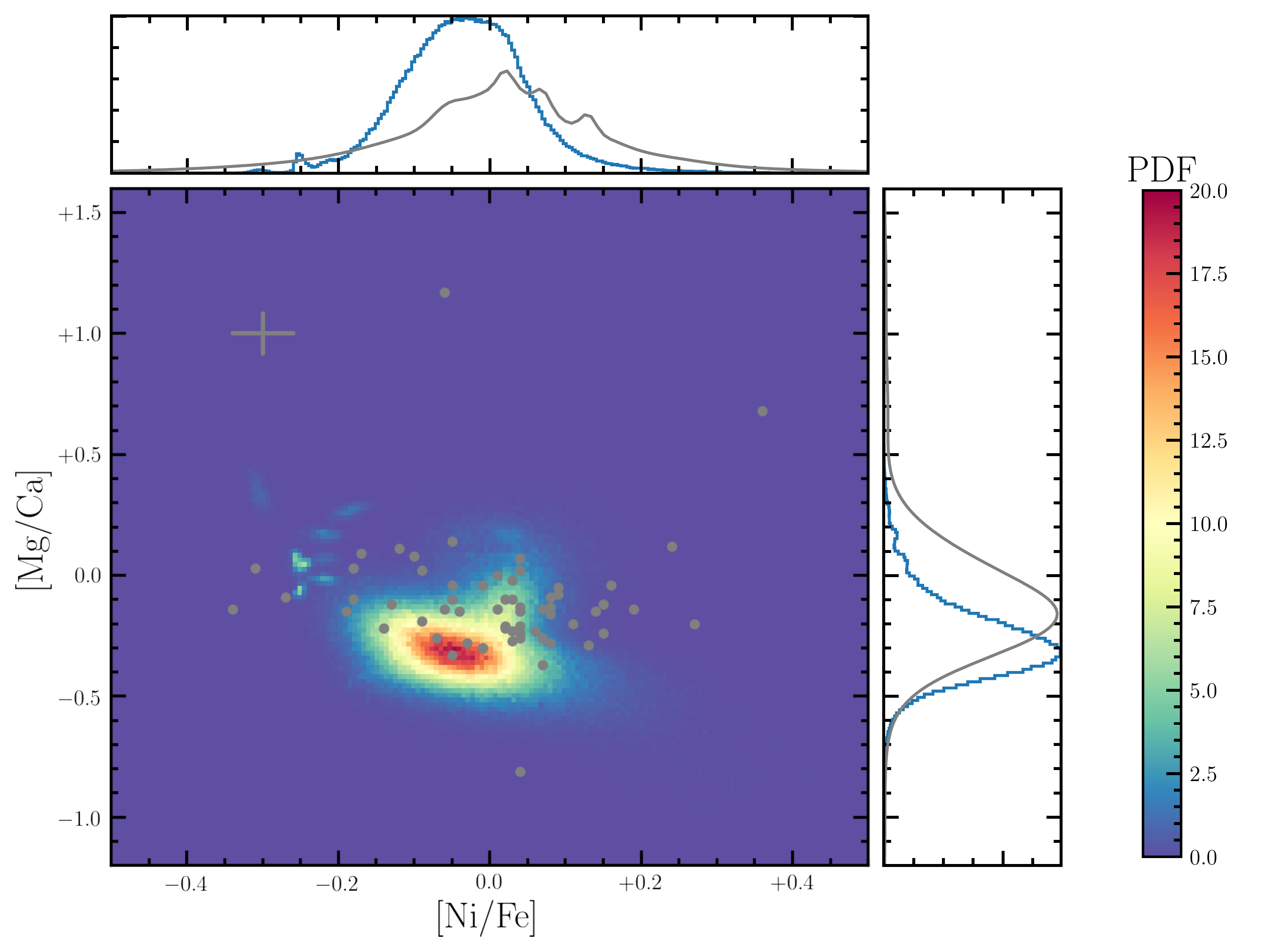}
    \caption{Same as Figure~\ref{fig:model_v_data_all} for the case when we exclude the \protect\citetalias{Roederer2014} data from our sample.}
    \label{fig:model_v_data_sub}
\end{figure*}

\indent To investigate the quality of the estimated enrichment models, we use the posterior distributions of the model parameters to generate the expected stellar abundances of [Mg/Ca] and [Ni/Fe]; we then compare these distributions to the observed abundances of the stellar sample. This  exercise indicates that our inferred model parameters, based on the \citetalias{Cayrel2004}, \citetalias{Bonifacio2009}, \citetalias{Yong2013}, and \citetalias{Roederer2014} samples, does not encompass the full extent of the data. In particular, Figure~\ref{fig:model_v_data_all} highlights a bimodal distribution of [Ni/Fe], which is a result of the bimodal distribution of $E_{\rm exp}$ (and hence $N_{\star}$) seen in Figure~\ref{fig:corner}. As can be seen in Figure~\ref{fig:yields_2x2_exp}, there are a range of models that can capture the [Ni/Fe] range seen in the data ($-0.4 \lesssim$~[Ni/Fe]~$ \lesssim +0.4$), including minimally mixed weak SNe and high energy SNe with `normal' mixing. However, these two possible ways to explain the broad [Ni/Fe] distribution are unable to simultaneously reproduce the observed [Mg/Ca] distributions. In particular, our model does not simultaneously favour supersolar [Mg/Ca] and [Ni/Fe] abundances. This highlights the importance of simultaneously modelling the interdependence of the $p(R_{i} | R_{m})$ distributions for all elements being considered in a stochastic enrichment model, as discussed in Section~\ref{sec:sampling}. \\
\indent From Figure~\ref{fig:model_v_data_sub}, we can see that the distinction between the model and observed distributions are less pronounced when we repeat our analysis after removing the \citetalias{Roederer2014} data. The \citetalias{Roederer2014} data comprise most of our sample, and exhibit somewhat elevated [Mg/Ca] values, compared with the rest of the sample (see Figure~\ref{fig:raw_data}). While this offset is $\sim0.1$~dex, combined with the broad distribution of [Ni/Fe], it is large enough to affect the parameter constraints of our model. To investigate this further, we have compared the abundances of the stars that appear in both the {\it First Stars} series and the \citetalias{Roederer2014} sample (in total, there are 26 duplicate stars that also meet our selection criteria). For these stars, the median [Mg/Ca] offset is $0.14 \pm 0.07$ and the median [Ni/Fe] offset is $0.05 \pm 0.13$ (where the quoted confidence interval represents a robust estimate of the standard deviation). We can use these offsets to apply a blanket `correction' to the \citetalias{Roederer2014} sample. Repeating our analysis, given this modified sample, produces parameter estimates consistent with those found after removing the \citetalias{Roederer2014} data (see Appendix~\ref{sec:mod_corner} for the associated corner plot).
We therefore conclude that our stochastic chemical enrichment model is able to reproduce the observations if we exclude the \citetalias{Roederer2014} data (or, indeed, calibrate these data to that of the other samples). As discussed in Section~\ref{sec:data}, the somewhat elevated Mg abundances reported by \citetalias{Roederer2014} are due to their adopted approach for estimating the effective temperature. 
Going forward, it is clear that the relative element abundance measurements need to be reported with an accuracy of better than $\sim0.05$~dex. \\
\indent Given that our model is able to reproduce the observations when we exclude the \citetalias{Roederer2014} data (see Figure~\ref{fig:model_v_data_sub}), we favour the maximum likelihood results of our model parameters based on the \citetalias{Cayrel2004}, \citetalias{Bonifacio2009}, and \citetalias{Yong2013} combined sample. We find that the average number of massive Population III stars that best describe the observed [Mg/Ca] and [Ni/Fe] abundances of these Population II stars is $\hat{N}_{\star} = 5^{+13}_{-3}$ where the quoted errors, here and subsequently, are the 68 per cent confidence region associated with the maximum likelihood value. These stars form obeying a power-law IMF with a slope
$\hat{\alpha} =2.9^{+0.9}_{-0.6}$. The maximum mass of the enriching stars is unconstrained while the marginalised 1D posterior distribution of $E_{\rm exp}$ tends towards the two extremes; if we consider the scenario with $N_{\star} <18$, then we infer $\hat{E}_{\rm exp} > 6~{\rm B}$~(95 per cent confidence). Finally, the preferred mixing prescription is  $\hat{f}_{\rm He} =  0.03^{+0.10}_{-0.03}$. Thus, our analysis suggests that the observed abundances of this metal-poor stellar sample are best described by enrichment from a small handful of Pop III hypernovae whose progenitors experienced minimal mixing between the stellar layers. Additionally, the slope of the underlying IMF (at least for masses that exceed 10~M$_{\odot}$) is consistent with that of a Salpeter distribution ($\alpha = 2.35)$.

\section{Discussion}
\label{sec:disc}
We first investigate the sensitivity of our results to the stellar atmosphere modelling. We then discuss the limitations of our model, and in particular, the chemical yields of massive stars. We close this section by drawing comparisons between the analysis of the most metal-poor stars and DLAs. \\
\subsection{Departures from local thermodynamic equilibrium}
\label{sec:nlte}
One of the assumptions that underpins the observational data is that the absorption lines are all formed in regions of local thermodynamic equilibrium (LTE). Neglecting departures from LTE is common practice when determining the stellar chemical abundances. Including non-LTE processes requires computationally expensive calculations that are generally dependent on the metallicity and surface temperature of the star under consideration, and requires knowledge of the radiative and collisional processes that drive the gas out of LTE. These calculations are unique to the element being considered and, indeed, are also influenced by its initial abundance \citep{Andrievsky2010}. 
While difficult to compute, these calculations can improve both the accuracy and the precision of the Mg, Ca, and Ni abundance determinations. \\
\indent Non-LTE Mg abundances for a subset of our stellar sample have been computed by \citet{Andrievsky2010}. 
The non-LTE corrections to Mg typically increase the Mg abundance by $\sim0.3$~dex.  Although non-LTE corrections to Ca are not currently available for the stars considered in this work, we note that the relative correction to the [Mg/Ca] abundances depends on metallicity \citep{Ezzeddine2018}; around [Fe/H]~$\simeq -3$, the [Mg/Ca] abundance should be reduced by $\sim0.1$~dex. Furthermore, this correction becomes more significant at even lower metallicity \citep{Sitnova2019}. Therefore, applying a non-LTE correction to Mg and Ca might bring our model into better agreement with the data. We look forward to a more detailed assessment of the [Mg/Ca] abundances when non-LTE corrections are available for a large fraction of the stellar sample employed in this work.

\subsection{Assessment of the yields}
We now consider potential improvements that could be made to the nucleosynthetic yields used in our model. As discussed in the previous section, the yields are unable to produce supersolar [Mg/Ca] at the same time as supersolar [Ni/Fe] (see Figures~\ref{fig:yields_2x2_exp} and \ref{fig:model_v_data_all}). When removing the \citetalias{Roederer2014} data (see Figure~\ref{fig:model_v_data_sub}), the model cannot reproduce the large spread of [Ni/Fe] values; in particular, the highest values of [Ni/Fe] deviate from the best-fit model.
Assuming that the observed abundances are reliable, this suggests that either: (1) massive metal-free stars are not the only sources responsible for the chemical enrichment of the most metal-poor stars; or (2) the simulated yields do not fully capture the physics that is necessary to reproduce the observed abundances.  \\ 
\indent Considering the first possibility, there are a number of alternative sources of enrichment that are not considered in this work. These include low mass metal-free stars (e.g \citealt{CampbellLattanzio2008}),  pair instability SNe (PISNe) (e.g. \citealt{HegerWoosley2002}), pulsational pair-instability SNe (pPISNe) (e.g. \citealt{Woosley2017}), and rapidly-rotating near-pristine massive stars (e.g. \citealt{Meynet2010}). There are reasonable reasons to discount the first three sources, for example:
(1) the most metal-poor stars are believed to have formed in the very early universe, when there would not have been enough time for low and intermediate mass stars to contribute their enrichment.
(2) PISNe produce a lot of metals, but their distinct chemical signatures (e.g. a low [$\alpha$/Fe] ratio, and a strong odd-even effect) are not seen in any of the stars of our sample. Furthermore, the yields of PISNe are expected to be incorporated in stars of metallicity [Fe/H]~$\sim -2.0$ (e.g. \citealt{Aoki2014}) which is somewhat higher than the stars in our sample.
(3) pPISNe also produce a distinct chemical signature (including a very high [$\alpha$/Fe] ratio). This signature is not observed in any stars of our sample. Finally, although we cannot discount enrichment from the rotating near-pristine stars  based on the current data, this may become possible in the future by exploring a larger grid of models or by measuring the helium mass fraction of the stellar sample. \\
\indent We now consider the possibility that the simulated yields are not yet able to fully capture the physics underpinning the stellar evolution and SN explosions of metal-free stars. In recent years substantial progress has been made in simulating CCSNe in 3D (e.g. \citealt{Vartanyan2018}). However, the mechanism that drives CCSNe is still unknown. We therefore lack a description of these SNe from first principles, and thus model calculations need to parameterise the explosion model. As discussed in Section~\ref{sec:model}, the \citetalias{HegerWoosley2010} simulations are performed in 1D using non-rotating models. Calculations performed in multiple dimensions are better able to capture the impact of both Rayleigh-Taylor instabilities and stellar rotation \citep{Joggerst2010-ROT, Joggerst2010-R-T, Vartanyan2018}. Furthermore, multidimensional models allow for departures from spherical symmetry, providing a more physically motivated scenario. We refer the reader to \citet{Muller2019} for a discussion of potential ways to observationally decipher this explosion mechanism and to \citet{Muller2020} for a review of the state-of-the-art simulations in this field. \\
\indent Indeed, being able to accurately simulate the complexity of a CCSN is a tall order. Surveys of metal-poor stars (e.g. \citealt{Starkenburg2017, DaCosta2019}) are uncovering a slew of chemically peculiar stars whose abundances are challenging to explain through the yields of CCSNe alone. An analysis of the abundances of UMP stars has shown that the \citetalias{HegerWoosley2010} yields are not always sufficient \citep{Placco2016}. Further, the recent detection of elevated Zn in the chemically peculiar star HE~1327-2326 has motivated the consideration of aspherical SNe models \citep{Ezzeddine2019}.
We thus conclude that nucleosynthetic yields provide an illustrative model of chemical enrichment, but because of the various simplifications involved, it is not yet clear how accurately the yields will represent the data. Moving forward, we highlight the importance of quantifying the errors of nucleosynthetic yields, and including this uncertainty in the modelling of observational data.  \\

\subsection{Comparison with DLAs}
A complementary approach to studying the chemistry of stellar relics is the study of minimally processed gas at high redshift \citep{Erni2006, Pettini2008, Penprase2010}. There are some gaseous systems at $z\sim3-4$ that appear to show \emph{no} discernible metals (e.g. \citealt{Fumagalli2012} and \citealt{Robert2019}). These systems may have remained \emph{entirely} untouched by the process of star formation. There are other systems, whose metallicities are comparable to that of EMP stars, that may have been solely enriched by the first generation of stars (e.g. \citealt{Cooke2017}). These systems, defined as DLAs if the column density of neutral hydrogen exceeds $\log_{10}\,N(\rm H\,${\sc i}$)/\rm cm^{-2}>20.3$, offer an alternative environment to search for a unique Population III signature. \\
\indent In contrast to stellar relics, whose abundance determinations require the consideration of complex processes, the physics required to determine the chemical composition of DLAs is rather simple; the column density of neutral hydrogen in these systems is sufficiently high to self-shield the gas, leaving all metals in a single dominant ionisation state. We can therefore determine the elemental abundances of these systems with a high degree of precision ($\sim 0.01$~dex; \citealt{Wolfe2005}). 
Additionally, when investigating these systems (as done in \citealt{Welsh2019, Welsh2020}) we can utilise the abundance ratios of the most abundant chemical elements, including [C/O]. The simulated [C/O] \citetalias{HegerWoosley2010} yields share an almost monotonically decreasing relationship with progenitor star mass. This is invaluable when attempting to estimate the mass distribution of the enriching stars. 
While the abundances of DLAs are, in principle, more straightforward to determine, these systems do not necessarily probe the multiplicity of the first stars (i.e. the average number of massive Population III stars forming in a given minihalo). Instead, it is possible that the constituent DLA gas has originated from multiple minihaloes. 
\begin{figure}
    \centering
    \includegraphics[width = \columnwidth]{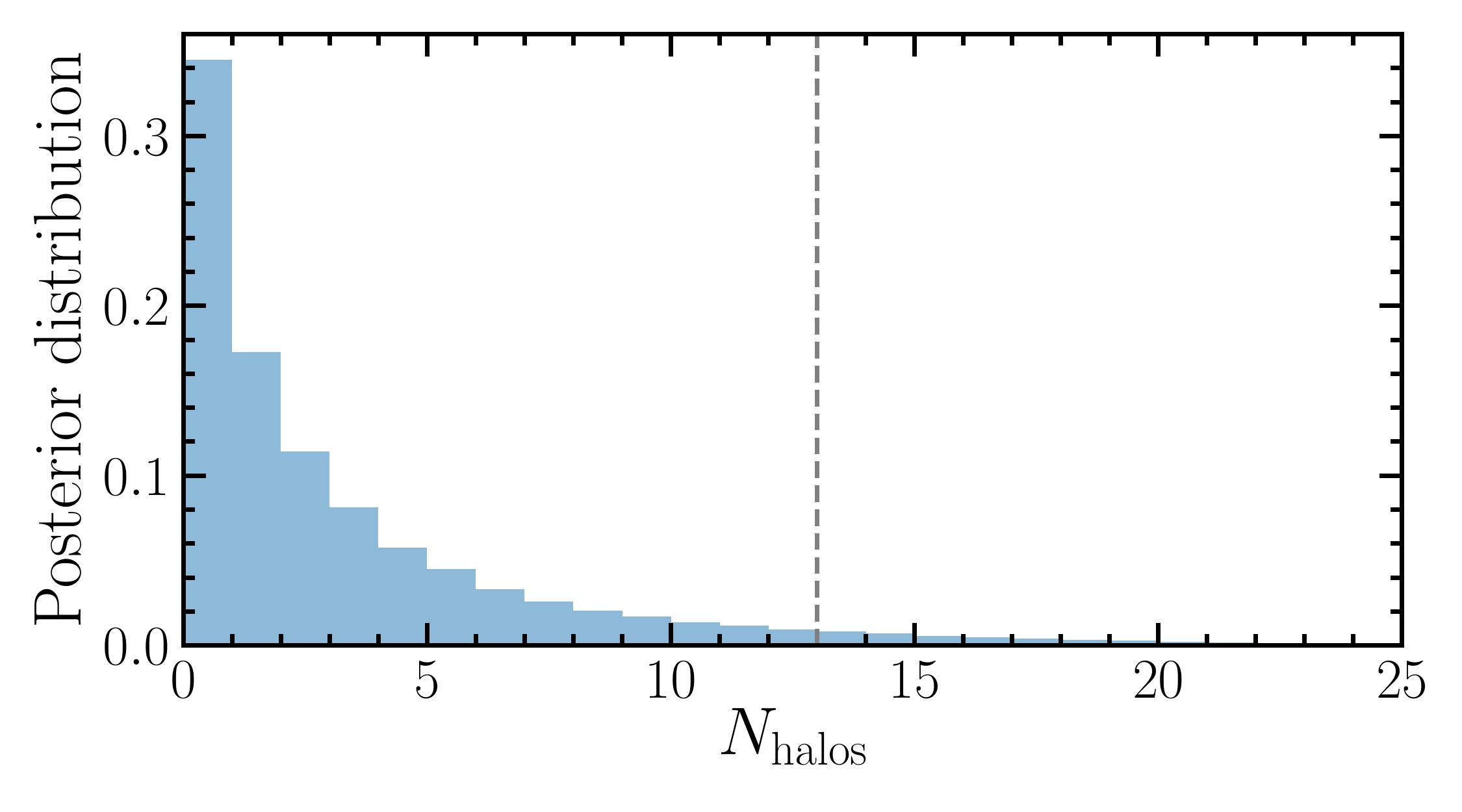}
    \caption{The posterior distribution of the average number of minihaloes that may have chemically contributed to the most metal-poor DLAs given our estimate of $N_{\star \, \rm DLAs}$ from \protect\citealt{Welsh2019} and our estimate of $N_{\star \, \rm Pop \, II}$ from this work. The vertical dashed line indicates the 95$^{\rm th}$ percentile of this distribution and corresponds to our quoted $2\sigma$ upper limit. }
     \label{fig:N_halo}
\end{figure}
Thus, the combined analysis of both EMP DLAs and EMP stars offers a novel opportunity to investigate the number of minihaloes that have chemically contributed to these high redshift structures (i.e. $N_{\rm halos} = N_{\star \,\rm DLAs}/ N_{\star \, \rm Pop \, II}$). Given the currently available data and suite of yield calculations, we can estimate the posterior distribution of $N_{\rm halos}$ given the inference on $N_{\star \,\rm DLAs}$ from \citet{Welsh2019} and $N_{\star \, \rm Pop \, II}$ from this work. The results of this comparison are shown in Figure~\ref{fig:N_halo} and indicate that the maximum likelihood value of $\hat{N}_{\rm halos}  = 1$. The most metal-poor DLAs may therefore contain the chemical products from just 1 or 2 minihaloes; they could also represent the clouds of gas from which some Population II stars formed. However, the tail of the $N_{\star \,\rm DLAs}$ distribution is quite broad and we therefore conservatively conclude $N_{\rm halos} < 13$ (95 per cent confidence). This upper limit is indicated by the vertical dashed line in Figure~\ref{fig:N_halo}. Given more precise constraints, this tool may provide a test of galaxy formation on the smallest scales. We look forward to comparing the enrichment histories of these systems in detail with future data.

\section{Conclusions}
\label{sec:conc}
In this paper we have applied a novel stochastic chemical enrichment model to investigate the possible enrichment history of a sample of metal-poor Milky Way halo stars using the [Mg/Ca] and [Ni/Fe] abundances from four historic surveys (\citetalias{Cayrel2004}, \citetalias{Bonifacio2009}, \citetalias{Yong2013}, and \citetalias{Roederer2014}). This is the first analysis to consider the number of massive Population III stars that have chemically enriched these stellar relics. Our main conclusions are as follows:
\begin{enumerate}
    \item With the adopted nucleosynthetic yields, our stochastic chemical enrichment model is able to reproduce the observed abundances of a stellar subsample comprising the \citetalias{Cayrel2004}, \citetalias{Bonifacio2009}, and \citetalias{Yong2013} data. In this scenario, our model shows preferential enrichment from a low number of Population III hypernovae.
    \smallskip
     \item Specifically, this model suggests that a typical metal-poor star has been chemically enriched by  $\hat{N}_{\star\,} = 5^{+13}_{-3}$ Population III hypernovae ($\hat{E}_{\rm exp} > 6~{\rm B}$~; 95 per cent confidence) that experienced minimal mixing between the stellar layers ($\hat{f}_{\rm He} =  0.03^{+0.10}_{-0.03}$). The IMF slope of this enriching population is found to be broadly consistent with that of a Salpeter distribution ($\hat{\alpha} =2.9^{+0.9}_{-0.6}$). Unless otherwise stated, these errors represent the 68 per cent confidence intervals of our estimates.
     \smallskip
    \item We consider the value of $N_{\star}$ reported in this paper to be a proxy for the multiplicity of massive Population III stars. By comparing this result to the $N_{\star}$ value inferred for the most metal-poor DLAs (see \citealt{Welsh2019}), we estimate the average number of minihaloes that may have chemically contributed to these low mass structures at $z\sim2-3$. We find $N_{\rm halos} < 13$ (95 per cent confidence). The maximum likelihood value of this distribution suggests that the most metal-poor DLAs may contain the chemical products from only a few minihaloes. In future, with more precise constraints, we hope to probe whether these DLAs resemble the gas that some Population II stars formed from.
    \smallskip
    \item In our analysis we utilise the abundance ratios calculated under the assumption of LTE. Recent work considering the non-LTE corrections to the observed [Mg/Ca] abundance ratios of these data may bring our model into better agreement with the data \citep{Ezzeddine2018, Sitnova2019}. Similarly, the different methods used to infer the effective temperature of the stars in our sample produce a [Mg/Ca] offset of $\sim +0.1$~dex (see the discussion in \citetalias{Roederer2014}). Although this difference is small, we find that the yields (and our model parameters) are sensitive to these small changes in abundance ratios. Thus, in future analyses, we require abundance measurements reported with an accuracy of $\sim 0.05$~dex to produce reliable estimates.
    \smallskip
    \item We also comment that the various parameterisations involved in nucleosynthetic yield calculations mean that it is not yet clear how accurately the adopted yields capture all of the relevant physics. Going forward, it would be helpful to consider the uncertainties of the simulated yields in future analyses.
\end{enumerate}
Finally, we emphasis that utilising this model to investigate the enrichment of both Population II stars and the \emph{most} metal-poor DLAs may reveal, not only the multiplicity of the first stars, but the chemical enrichment of some of the lowest mass structures at $z\sim 3$. It is therefore a promising tool for investigating early structure formation. Future modifications to this model may offer an alternative way to study inhomogeneous metal-mixing and the possibility of externally enriched minihaloes.

\section*{Acknowledgements}
During this work, R.~J.~C. was supported by a
Royal Society University Research Fellowship.
We acknowledge support from STFC (ST/L00075X/1, ST/P000541/1).
This project has received funding from the European Research Council 
(ERC) under the European Union's Horizon 2020 research and innovation 
programme (grant agreement No 757535).
This work used the DiRAC Data Centric system at Durham University,
operated by the Institute for Computational Cosmology on behalf of the
STFC DiRAC HPC Facility (www.dirac.ac.uk). This equipment was funded
by BIS National E-infrastructure capital grant ST/K00042X/1, STFC capital
grant ST/H008519/1, and STFC DiRAC Operations grant ST/K003267/1
and Durham University. DiRAC is part of the National E-Infrastructure.
This research has made use of NASA's Astrophysics Data System.
\section*{Data availability}
No new data were generated or analysed in support of this research.


\bibliographystyle{mnras}
\bibliography{references} 


\appendix
\section{Intrinsic scatter calculation}
\label{sec:append}
\begin{figure*}
    \centering
    \includegraphics[width=\textwidth]{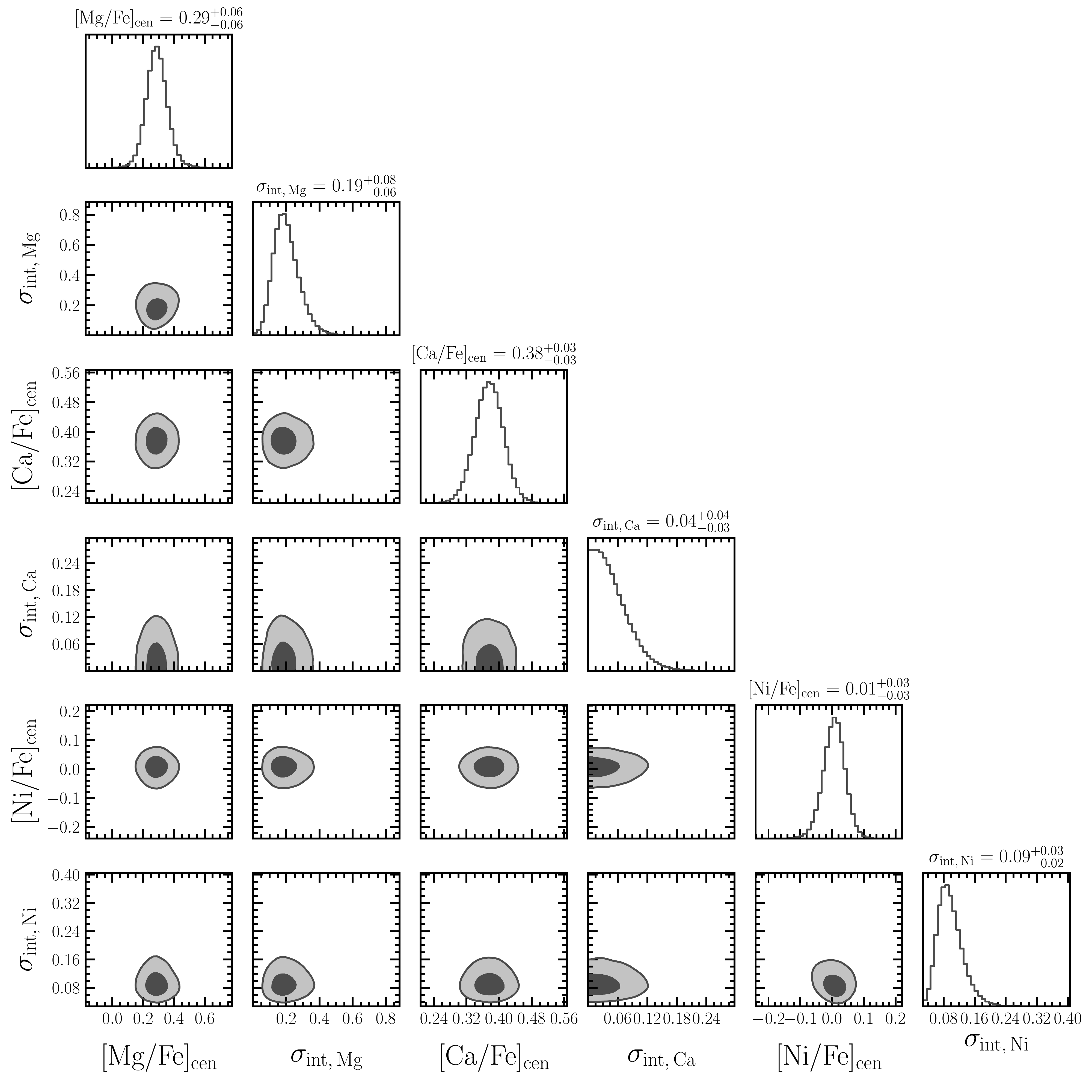}
    \caption{Converged MCMC analysis of the intrinsic scatter of the \protect\citetalias{Cayrel2004} data. The main diagonal panels show the marginalised maximum likelihood distributions of the parameters describing the central values and intrinsic dispersions of the \protect\citetalias{Cayrel2004} data (refer to Section~\ref{sec:scatter}). The associated contours highlight their associated 2D projections. The dark and light contours show the $68\%$ and $95\%$ confidence regions of these projections, respectively. }
    \label{fig:cay}
\end{figure*}

In this appendix, we describe our approach to estimate the intrinsic scatter of the stellar samples. This scatter is defined by Eq.~\ref{eqn:sig}. As described in Section~\ref{sec:scatter}, we consider two model parameters ([X/Fe]$_{\rm cent}$ and $\sigma_{\rm int}$) that we determine independently for each abundance ratio and each sample. We use an MCMC procedure to simultaneously estimate the central values [Mg/Fe]$_{\rm cent}$, [Ca/Fe]$_{\rm cent}$, and [Ni/Fe]$_{\rm cent}$ of a sample, alongside their associated additional error components $\sigma_{\rm int, Mg}$, $\sigma_{\rm int, Ca}$, and $\sigma_{\rm int, Ni}$. We adopt uniform priors for these parameters; the central values are bounded by $-5 \leq [\rm X/Fe]_{\rm cent} \leq 5$ and the intrinsic errors are bounded by $0 \leq \sigma_{\rm int, X} \leq 5 $. We utilised the {\sc emcee} software package \citep{EMCEE} to randomly initialise 400 walkers and explore this parameter space. The results of this analysis for the \citetalias{Cayrel2004} data are presented in Figure~\ref{fig:cay} which shows the converged posterior distributions of these model parameters. The confidence interval for [Mg/Fe] demonstrates there is a statistically significant deviation from 0; this result is replicated across all of the stellar samples. We consider the possibility that this intrinsic dispersion is the result of stochastic sampling of the IMF.

\section{Analysis of the corrected R14 sample}
\label{sec:mod_corner}
\begin{figure*}
    \centering
    \includegraphics[width=\textwidth]{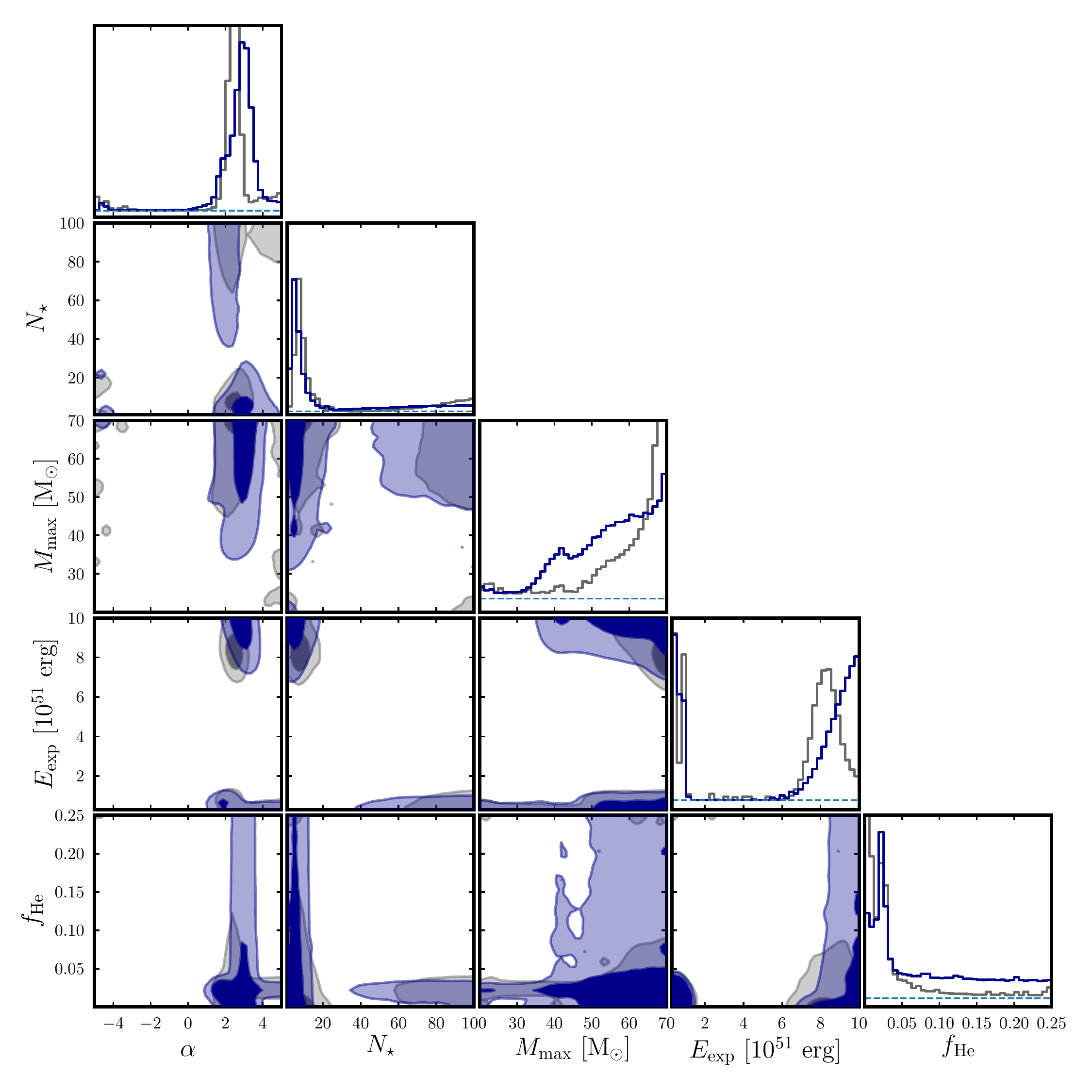}
    \caption{Same as Figure~\ref{fig:corner} using the offset corrected \protect\citetalias{Roederer2014} [Mg/Ca] and [Ni/Fe] data in place of the original values.}
    \label{fig:mod_corner}
\end{figure*}

In this appendix, we show the results of repeating our analysis using the offset corrected \citetalias{Roederer2014} data in combination with the \citetalias{Cayrel2004}, \citetalias{Bonifacio2009}, and \citetalias{Yong2013} data. We use the abundances of the stars that appear in multiple surveys to calculate the [Mg/Ca] and [Ni/Fe] offsets of $0.14 \pm 0.07$ and $0.05\pm0.13$ respectively. The confidence intervals are estimated using the median absolute deviation; for normally distributed data, $\sigma\simeq1.4826\,{\rm MAD}$. 
When applying this correction, the errors associated with the \citetalias{Roederer2014} [Mg/Ca] and [Ni/Fe] abundances are given by the reported observational errors and these additional systematic components added in quadrature. 
Figure \ref{fig:mod_corner} shows the parameter estimates that result from considering this modified sample (grey distributions). Unlike those produced using the original data (see Figure~\ref{fig:corner}), these estimates are consistent with those found using the reduced sample (blue distributions in Figure~\ref{fig:mod_corner}). Given that we have calibrated the \citetalias{Roederer2014} data using a substantial portion of the stars within the reduced sample, this agreement is reassuring. We note that the original discrepancies between the sample distributions are invaluable for testing the sensitivity of our model to the input data. 

\bsp	
\label{lastpage}
\end{document}